\documentclass[aps,prb,reprint,longbibliography,amsmath,amssymb,floatfix]{revtex4-2}
\usepackage{graphicx}
\usepackage{bm}
\usepackage{float}
\usepackage[colorlinks=true,citecolor=blue,linkcolor=blue,urlcolor=blue]{hyperref}
\hypersetup{pdftitle={Surface criticality in the low-energy entanglement spectrum of the transverse-field Ising model}}

\makeatletter
\AtBeginDocument{\immediate\write\@auxout{\string\citation{apsrev42Control}}}
\makeatother
\begin{document}
\title{Surface criticality in the low-energy entanglement spectrum\\of the (2+1)D transverse-field Ising model}
\author{Jiwen Chen}
\affiliation{Department of Physics, School of Science and Research Center for Industries of the Future, Westlake University, Hangzhou 310030, China}
\affiliation{Institute of Natural Sciences, Westlake Institute for Advanced Study, Hangzhou 310024, China}

\author{Zhe Wang}
\email{wangzhe90@westlake.edu.cn}
\affiliation{Department of Physics, School of Science and Research Center for Industries of the Future, Westlake University, Hangzhou 310030, China}
\affiliation{Institute of Natural Sciences, Westlake Institute for Advanced Study, Hangzhou 310024, China}

\author{Zheng Yan}
\email{zhengyan@westlake.edu.cn}
\affiliation{Department of Physics, School of Science and Research Center for Industries of the Future, Westlake University, Hangzhou 310030, China}
\affiliation{Institute of Natural Sciences, Westlake Institute for Advanced Study, Hangzhou 310024, China}

\author{Bin-Bin Mao}
\email{maobinbin@uhrs.edu.cn}
\affiliation{School of Foundational Education, University of Health and Rehabilitation Sciences, Qingdao 266000, China}

\begin{abstract}

How is physical surface criticality encoded in the low-energy entanglement spectrum? Here we show that the entanglement spectrum provides a way to separate the surface degrees of freedom from the bulk, revealing distinct physical mechanisms underlying different surface critical regimes. We study the (2+1)-dimensional transverse-field Ising model at its bulk quantum critical point, taking a complete physical surface chain as the subsystem. By combining the reduced density matrix sampling and the multi-replica tricks in quantum Monte Carlo, we investigate the low-energy spectra and magnetic responses of the corresponding entanglement Hamiltonian. In the ordinary regime, the low-energy spectrum exhibits finite gaps in the probed momentum sectors, consistent with a surface behavior governed by bulk critical fluctuations. In the extraordinary regime, an increasingly near-degenerate zero-momentum doublet emerges, reflecting the spectral signature expected from surface ordering. Most intriguingly, near the special coupling, both low-energy gaps display similar power-law scaling, with effective exponents close to 1/2. This behavior points to a nontrivial interplay between bulk critical fluctuations and gapless surface degrees of freedom. Our results demonstrate that the entanglement spectrum does more than characterize the surface excitations: it offers a distinct perspective on how bulk and surface critical modes organize and interact across different surface critical regimes.
\end{abstract}
\maketitle
\raggedbottom

\section{Introduction}
\label{sec:introduction}
The universal behavior of a critical system is not fully characterized by its bulk universality class when physical boundaries are present. At criticality, boundary degrees of freedom can exhibit scaling dimensions~\cite{Hu2026scaling,Cardy1984Conformal,Dedushenko2024Ising,Gliozzi2015,Francesco2021}, correlation functions~\cite{Przetakiewicz2025Boundary,Hu2021correlation,ParisenToldin2022Boundary,Zhang2017Unconventional,Weber2018correlation}, and ordering properties~\cite{Deng2005Surface,Hasenbusch2011Monte,Ding2023Special,Metlitski2022Boundary,LvPRB2022} distinct from those in the bulk, governed by their own renormalization-group fixed points. Surface critical phenomena therefore provide an important framework for understanding how bulk critical fluctuations influence lower-dimensional boundary degrees of freedom and how physical boundaries enrich the universal behavior of critical systems~\cite{Diehl1997The,Binder1974Surface,Cardy1984Conformal,Zhang2017Unconventional}.

At a bulk quantum critical point, varying the surface interactions can realize distinct boundary critical regimes while leaving the bulk criticality unchanged~\cite{Ding2023Special,Shu2026Universal}. In the presence of a physical boundary, the surface critical behavior is governed by boundary renormalization-group fixed points. For an Ising order parameter, weak surface enhancement leads to ordinary surface criticality, sufficiently strong enhancement supports an ordered extraordinary surface, and the special point separates these regimes~\cite{Shu2026Universal,Hasenbusch2011Monte,Deng2005Surface}. These boundary fixed points exhibit distinct surface correlations and scaling properties, providing a controlled setting for investigating how different boundary universality classes are reflected in quantum entanglement.

Retaining a complete physical surface and tracing out all remaining degrees of freedom defines a surface reduced density matrix (RDM) $\rho_A$ and an entanglement Hamiltonian $H_E=-\ln\rho_A$~\cite{Dalmonte2022Entanglement,Li2024Relevant,Wang2025Sudden,Liu2026Worldline,Yang2026Limits}. The low-lying entanglement levels correspond to the largest eigenvalues of $\rho_A$, revealing level spacings and near-degeneracies beyond the information contained in the entanglement entropy~\cite{Li2008Entanglement,Peschel2009Reduced,Dalmonte2022Entanglement}. Although $H_E$ acts on a boundary degree of freedom, it incorporates the influence of the critical bulk~\cite{Cirac2011Entanglement}. 
This raises a central question: how do distinct surface critical regimes manifest themselves in the low-energy entanglement spectrum, and to what extent do their spectral features reflect the underlying boundary criticality?

Previous studies have connected surface criticality to entanglement spectra and entanglement ensemble in surface-related geometries~\cite{Zhu2026Bipartite,Zhu2026Criticality}. Motivated by these developments, we investigate the low-energy entanglement spectra of a single retained physical surface chain across the ordinary, special, and extraordinary regimes. We combine quantum Monte Carlo sampling of $\rho_A$~\cite{Mao2025Sampling,Mao2026Detecting,Wang2025Entanglement} with magnetic measurements in entanglement ensemble $\rho_A^n \propto e^{-H_E/T_E} $, corresponding to $T_E=1/n$ and $n$ is the number of replicas~\cite{Yan2023Unlocking,Li2024Relevant,Wang2025Sudden}. Momentum-resolved entanglement spectroscopy characterizes the low-energy entanglement level structure, while replica observables provide complementary information about magnetic ordering in the corresponding entanglement ensembles. Together, these approaches allow us to examine the relationship between low-energy spectral features and magnetic responses over different entanglement temperatures.

We study the ordinary, special, and extraordinary surface regimes of the square-lattice transverse-field Ising model at its bulk quantum critical point, where the physical surface behavior is governed by the 3D Ising boundary conformal theory~\cite{Dedushenko2024Ising,Zhou2025Studying}. By comparing the splitting between the two lowest zero-momentum levels with the gap to the nearest nonzero momentum, we identify distinct finite-size spectral structures: positive extrapolated gaps in the ordinary regime, approximately algebraic softening near the special coupling, and an increasingly near-degenerate zero-momentum pair in the extraordinary regime. Replica measurements in the entanglement ensemble at the same physical couplings reveal contrasting magnetic responses consistent with these spectral signatures. These results establish a connection between distinct surface critical regimes and their low-energy entanglement spectral structures, while highlighting the distinction and correspondence between physical boundary criticality and the low-energy properties of the entanglement Hamiltonian.

\begin{figure}[tbp]
 \centering
 \includegraphics[width=\columnwidth]{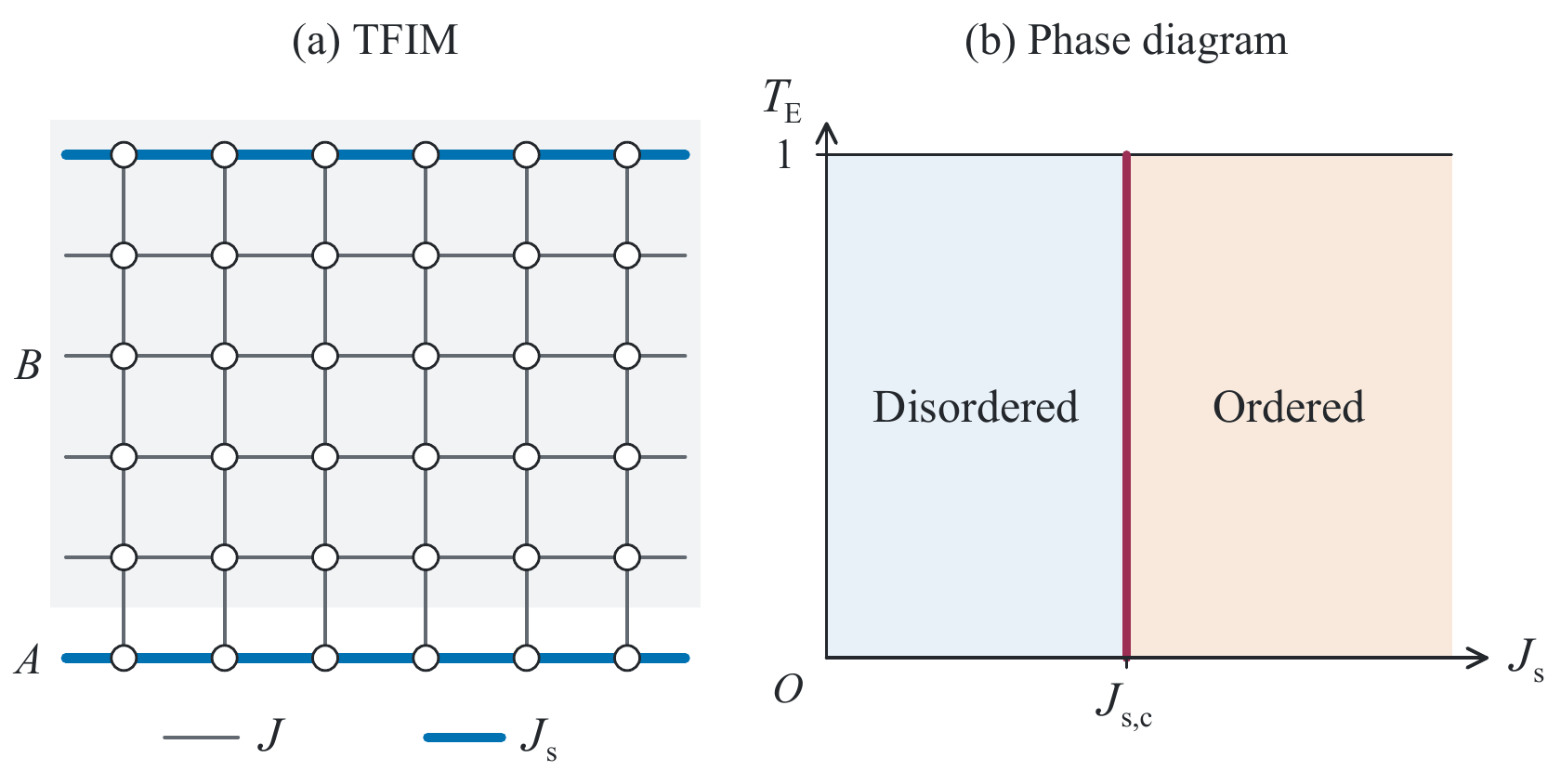}
 \caption{Schematic illustration of the lattice model and phase diagram. (a) Square-lattice TFIM with bulk bonds $J$ (gray) and tunable surface bonds $J_s$ (blue). The shaded region $B$ is traced out, leaving the bottom physical surface chain $A$ as the retained subsystem. Bond stubs indicate periodic continuation along $x$, while the $y$ direction has open boundaries. (b) Schematic phase diagram in the $(J_s,T_E)$ plane. The original physical reduced state corresponds to $T_E=1$, while replica calculations access $T_E=1/n$. The vertical boundary is illustrative.}
 \label{fig:setup}
\end{figure}

\section{Model and numerical methods}
\label{sec:model}

\subsection{Hamiltonian and surface geometry}
We consider the square-lattice transverse-field Ising model (TFIM),
\begin{align}
 H=-J\sum_{\langle ij\rangle\notin\mathcal S}\sigma_i^z\sigma_j^z
   -J_s\sum_{\langle ij\rangle\in\mathcal S}\sigma_i^z\sigma_j^z
   -h\sum_i\sigma_i^x,
 \label{eq:hamiltonian}
\end{align}
where $\sigma_i^{x,z}$ are Pauli matrices and $\mathcal S$ denotes the set of horizontal bonds along the two physical surfaces. We set $J=1$ as the energy unit, so that all other nearest-neighbor bonds, including those connecting the surfaces to the bulk, have unit strength. The lattice dimensions are $L_x=L_y=L$, with periodic boundary conditions along $x$ and open boundary conditions along $y$. Throughout this work, the transverse field is fixed at the bulk critical value $h_c=3.04438(2)$~\cite{Blote2002Cluster}.

We retain the complete physical surface chain ($y=0$) as subsystem $A$, while its complement $B$ consists of all remaining spins, including the opposite physical surface [Fig.~\ref{fig:setup}(a)]. The resulting RDM therefore describes the surface degrees of freedom while incorporating their coupling to the critical bulk.
We focus on three representative surface couplings, $J_s=1.0$, $1.9895$, and $3.0$, corresponding to the ordinary, special, and extraordinary physical surface regimes, respectively \cite{Ding2018Engineering}. The special reference coupling is chosen according to the estimate $J_s^*=1.9895(6)$ reported in Ref.~\onlinecite{Shu2026Universal}. 

\subsection{Surface reduced density matrix and spectral gaps}
At physical inverse-temperature $\beta$, the normalized RDM of subsystem $A$ and its associated entanglement Hamiltonian $H_E$ are defined as
\begin{align}
 \rho_A(\beta)
 &=\frac{\operatorname{Tr}_B e^{-\beta H}}
 {\operatorname{Tr}e^{-\beta H}},
 &H_E&=-\ln\rho_A.
 \label{eq:rdm}
\end{align}

\begin{figure}[tbp]
 \centering
 \includegraphics[width=\columnwidth]{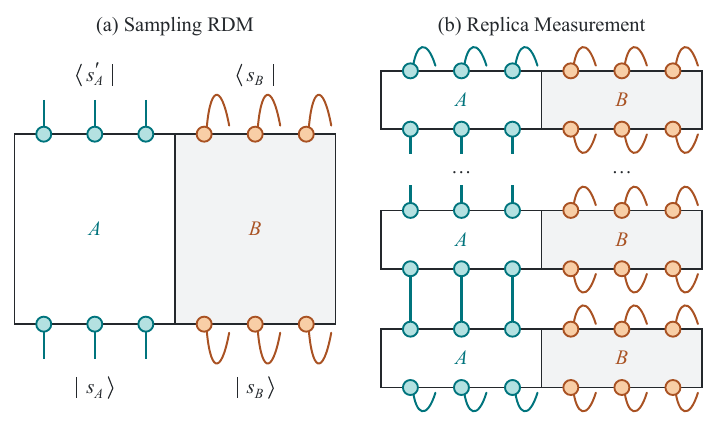}
 \caption{Entanglement-Hamiltonian interpretation of the imaginary-time manifolds.
 (a) Tracing out the environment $B$ defines the surface reduced density matrix $\rho_A=e^{-H_E}$. The retained physical surface $A$ is therefore described by an effective entanglement Hamiltonian $H_E$, which encodes the influence of the critical bulk on the surface degrees of freedom.
 (b) The $n$-replica construction realizes the normalized ensemble $\rho_A^n/\operatorname{Tr}\rho_A^n =e^{-H_E/T_E}/\operatorname{Tr}e^{-H_E/T_E}$, with $T_E=1/n$. Varying $n$ leaves $H_E$ unchanged and changes only its entanglement temperature, thereby progressively emphasizing its low-energy spectrum and associated magnetic response.
 }
 \label{fig:manifold}
\end{figure}

We obtain the surface RDM using direct quantum Monte Carlo (QMC) sampling, with open imaginary-time boundary conditions on $A$ and periodic imaginary-time boundary conditions on $B$~\cite{Mao2025Sampling}. For each QMC configuration in the imaginary-time manifold shown in Fig.~\ref{fig:manifold}(a), we record the spin configurations $|s_A\rangle$ and $\langle s'_A|$ at the lower and upper imaginary-time boundaries of $A$, respectively. The periodic closure of $B$ implements the trace over the environment.

Let $\mathcal W(s'_A,s_A)$ denote the statistical weight of QMC configurations with boundary pair $(s'_A,s_A)$. Its normalized sampling probability is
\begin{equation}
 P(s'_A,s_A)
 =\frac{\mathcal W(s'_A,s_A)}
 {\mathcal Z_{\mathrm{open}}},
 \qquad
 \mathcal Z_{\mathrm{open}}
 =\sum_{s'_A,s_A}\mathcal W(s'_A,s_A).
 \label{eq:open_probability}
\end{equation}
Since $\mathcal W(s'_A,s_A)$ is proportional to the corresponding matrix element of the unnormalized RDM $\operatorname{Tr}_B e^{-\beta H}$, the normalized RDM matrix elements can be reconstructed as
\begin{equation}
 \langle s'_A|\rho_A|s_A\rangle
 =\frac{\mathcal W(s'_A,s_A)}
 {\displaystyle\sum_{\tilde s_A}
 \mathcal W(\tilde s_A,\tilde s_A)}
 =\frac{P(s'_A,s_A)}
 {\displaystyle\sum_{\tilde s_A}
 P(\tilde s_A,\tilde s_A)}.
 \label{eq:rdm_sampling}
\end{equation}
The denominator enforces $\operatorname{Tr}_A\rho_A=1$. For the sign-free TFIM, all sampling weights are nonnegative, and $P(s'_A,s_A)$ can be estimated directly from the normalized Monte Carlo histogram of boundary configuration pairs.

The computational cost of this approach grows exponentially with the subsystem size. 
A surface chain of length $L$ has Hilbert space dimension $2^L$ and a full RDM with $4^L$ matrix elements, so resolving its low-lying levels demands ever more samples as $L$ grows. In practice, this confines direct entanglement spectroscopy to fairly small systems.

We obtain the entanglement spectra for $L=4,6,8,10,12$, accumulating $10^{10}$ Monte Carlo samples for each $(L,J_s)$, which is the typical sampling number for these system sizes~\cite{Mao2025Sampling,Mao2026Detecting}. The bulk is fixed at criticality with dynamical exponent $z=1$, therefore the physical inverse-temperature increases linearly with system size $\beta=2L$.



Because the reduced state inherits translation symmetry along the surface direction, it can be block-diagonalized into momentum sectors. Translation by one surface lattice site defines the allowed momenta
\begin{equation}
 k=\frac{2\pi\ell}{L},
 \qquad \ell=0,\ldots,L-1.
\end{equation}
Within each momentum sector, the positive eigenvalues $\lambda_j(k)$ of $\rho_A$ give the entanglement energies (entanglement Hamiltonian's eigenvalues)
\begin{equation}
 \xi_j(k)=-\ln\lambda_j(k),
 \label{eq:xi}
\end{equation}
ordered such that $\xi_0(k)\leq\xi_1(k)\leq\cdots$. For the ferromagnetic system considered here, the lowest level in the spectra lies at $k=0$. We denote its energy by $\xi_0\equiv\xi_0(0)$ and present the spectra relative to this reference level.

To characterize the low-energy spectral structures, we consider two excitation gaps,
\begin{align}
 \Delta_{q_1}(L)
 &=\xi_0(q_1)-\xi_0(0),
 &q_1&=\frac{2\pi}{L},
 \label{eq:gapk}\\
 \Delta_0(L)
 &=\xi_1(0)-\xi_0(0).
 \label{eq:gap0}
\end{align}
Here, $\Delta_0$ measures the splitting between the two lowest levels in the zero-momentum sector, whereas $\Delta_{q_1}$ probes the lowest excitation at the nearest nonzero momentum. Comparing their finite-size behavior allows us to distinguish the simultaneous softening of low-energy excitation scales from the formation of an increasingly near-degenerate zero-momentum pair. We will discuss the details in section \ref{sec:spectra}.

\subsection{Replica ensemble and magnetic observables}
To investigate the finite entanglement temperature properties of the surface entanglement Hamiltonian, we employ replica QMC. The unnormalized $n$-replicated state is defined as
\begin{equation}
\rho_A^n =e^{-H_E/T_E},
 \qquad T_E=\frac{1}{n},
 \label{eq:replica}
\end{equation}
where $H_E=-\ln\rho_A$. At fixed physical parameters and $\beta$, the entanglement Hamiltonian $H_E$ remains unchanged as the entanglement temperature $T_E$ is varied. Increasing the replica number therefore lowers $T_E$, allowing us to probe the thermal properties of the same entanglement Hamiltonian at different entanglement temperatures.

In the path-integral representation, the surface configurations are joined cyclically between replicas, while the imaginary-time boundaries of each copy of the environment $B$ are closed independently, as illustrated in Fig.~\ref{fig:manifold}(b)~\cite{Yan2023Unlocking,Hastings2010Measuring,Wang2025Sudden,Yang2026Limits}. This construction defines a partition function on the replicated manifold,
\begin{equation}
 \mathcal Z_{\mathrm{rep}}^{(n)}
 =\sum_{\mathcal C}\mathcal W_n(\mathcal C),
 \label{eq:replica_partition}
\end{equation}
where $\mathcal C$ denotes a complete QMC configuration and $\mathcal W_n(\mathcal C)$ is its statistical weight. With the unnormalized Boltzmann weight $e^{-\beta H}$ in each replica, the replicated partition function is related to the reduced density matrix by
\begin{equation}
 \begin{aligned}
 \mathcal Z_{\mathrm{rep}}^{(n)}
 &=\operatorname{Tr}_A
 \left[\left(\operatorname{Tr}_B e^{-\beta H}\right)^n\right]\\
 &=Z^n\operatorname{Tr}_A\rho_A^n\\
 &=Z^n\operatorname{Tr}_A e^{-H_E/T_E},
 \end{aligned}
 \label{eq:replica_partition_relation}
\end{equation}
where $Z=\operatorname{Tr}e^{-\beta H}$ is the partition function of the original system. The factor $Z^n$ is independent of the replica configuration and cancels in normalized expectation values. The corresponding normalized sampling probability is
\begin{equation}
 P_n(\mathcal C)
 =\frac{\mathcal W_n(\mathcal C)}
 {\mathcal Z_{\mathrm{rep}}^{(n)}},
 \qquad
 \sum_{\mathcal C}P_n(\mathcal C)=1.
 \label{eq:replica_probability}
\end{equation}

For a surface operator $O_A$ diagonal in the $\sigma^z$ basis, its value can be evaluated from the surface spin configuration at an imaginary-time boundary connecting adjacent replicas. Its expectation value in the replica ensemble is therefore the normalized QMC average
\begin{equation}
 \langle O_A\rangle_n
 =\operatorname{Tr}_A\left(
 \rho_A^{n}O_A\right)/\operatorname{Tr}_A\rho_A^{n}
 =\sum_{\mathcal C}P_n(\mathcal C)O_A(\mathcal C).
 \label{eq:replica_estimator}
\end{equation}
Here, $O_A(\mathcal C)$ denotes the value of the observable evaluated on the corresponding surface configuration. Owing to the cyclic symmetry of the replicated manifold, equivalent operator insertions can be made at any of the $n$ replica junctions. For the sign-free TFIM considered here, the sampling weights are nonnegative, so expectation values can be evaluated directly from the normalized Monte Carlo averages. Since the replica construction requires an integer number of copies, it provides direct access to the discrete entanglement temperatures $T_E=1/n$ for positive integer $n$.

An important property of this construction is that the $n=1$ ensemble $\mathcal Z_{\mathrm{rep}}^{(n=1)}$ exactly reproduces physical observables of the original system $Z(\beta)$. For any operator $O_A$ supported on the retained surface,
\begin{equation}
 \begin{aligned}
 \langle O_A\rangle_{\mathcal Z_{\mathrm{rep}}^{(n=1)}}
 =\operatorname{Tr}_A(\rho_A O_A)
 =\langle O_A\rangle_{Z(\beta)}.
 \end{aligned}
 \label{eq:identity}
\end{equation}
Thus, replica QMC at $n=1$ provides the physical surface observables at the physical inverse-temperature $\beta$, while $n>1$ accesses the thermal properties of the same entanglement Hamiltonian at lower entanglement temperatures.

We characterize the magnetic response of the retained surface using the magnetization per site,
\begin{equation}
 m=\frac{1}{2L}\sum_{x=0}^{L-1}
 \sigma^z_{x},
 \label{eq:magnetization}
\end{equation}
where $\sigma_x^z$ is the Pauli operator on site $x$ of the surface chain and the factor $1/2$ converts Pauli matrices to spin-$1/2$ operators. The second moment of magnetization measured at $T_E=1/n$ is 
\begin{equation}
\langle m^2\rangle_n = \operatorname{Tr}_A\left(
 \rho_A^{n}m^2\right)/\operatorname{Tr}_A\rho_A^{n},
 \label{eq:second-moment}
\end{equation}
and the fourth moment is defined in the similar way. We measure the second and fourth moments of the magnetization and construct the normalized Binder cumulant~\cite{Binder1981Finite}
\begin{equation}
 U_n=\frac32\left[
 1-\frac{\langle m^4\rangle_n}
 {3\langle m^2\rangle_n^2}
 \right].
 \label{eq:binder}
\end{equation}
This dimensionless cumulant vanishes for a zero-mean Gaussian magnetization distribution and approaches unity for an ideal symmetric distribution consisting of two sharp peaks at $m=\pm m_0$. Together, $\langle m^2\rangle_n$ and $U_n$ characterize the magnetic fluctuations and ordering tendencies of the surface in the replica ensemble.

We perform replica QMC simulations for $L=24,32,40$. At the representative surface couplings $J_s=1.0,1.9895,3.0$, we vary the replica number over $n=1,\ldots,9$. We additionally scan the surface coupling over $1.0\leq J_s\leq3.0$ at fixed replica numbers $n=1,2,6,10,20$. For the fixed-coupling and coupling-scan calculations, we perform forty and ten independent simulation runs, respectively, with typically $10^5$ measurements per run. We will discuss the details in section~\ref{sec:temperature}.

\begin{figure*}[tp]
 \centering
 \includegraphics[width=0.8\textwidth]{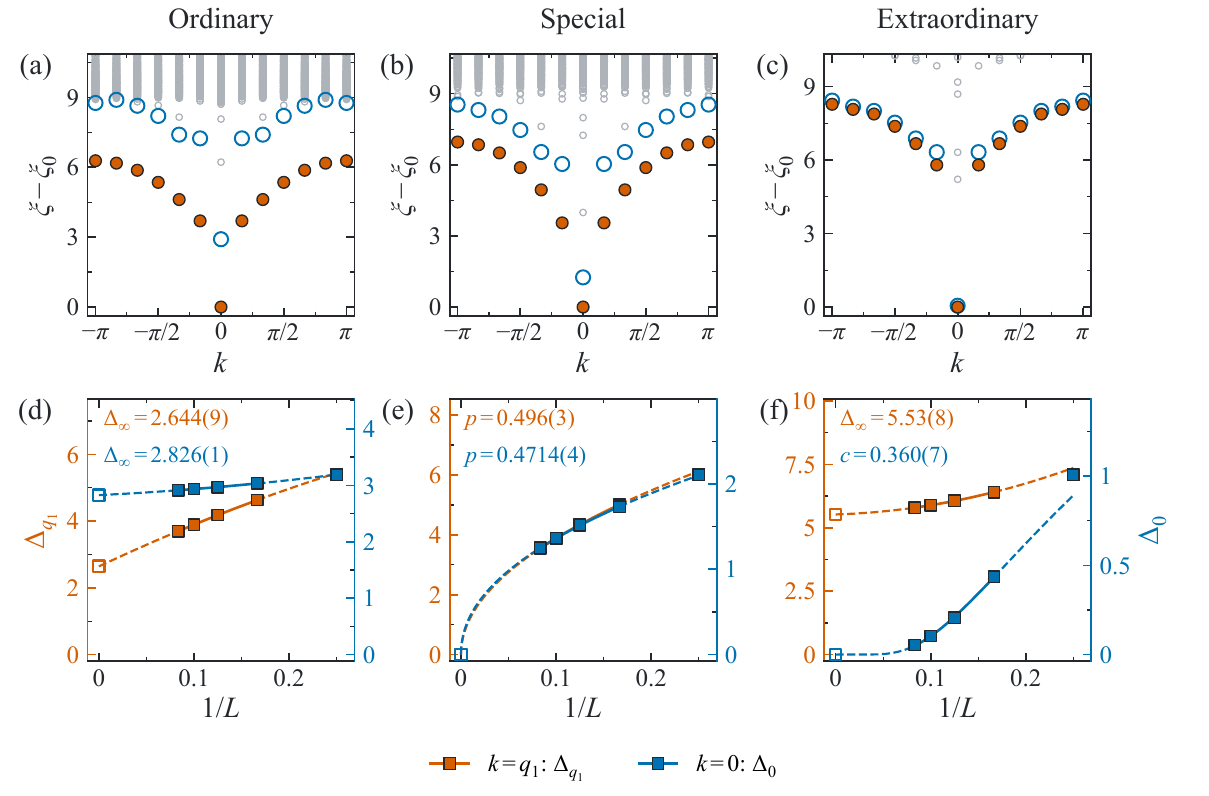}
 \caption{Surface entanglement spectra and finite-size gaps for ordinary ($J_s=1.0$), special ($J_s=1.9895$), and extraordinary ($J_s=3.0$) surfaces. (a)--(c) Spectra at $L=12$, shifted by $\xi_0$; positive eigenvalue estimates with $\xi\leq11$ are shown. Vermilion filled circles mark sector minima, blue open circles the second levels, and gray open circles higher levels. (d)--(f) $\Delta_{q_1}$ (vermilion filled squares, left axes) and $\Delta_0$ (blue filled squares, right axes). Fits use $L=6,8,10,12$; $L=4$ remains displayed. Both gaps in (d) and $\Delta_{q_1}$ in (f) use Eq.~\eqref{eq:finitegap}; both gaps in (e) use $AL^{-p}$, and $\Delta_0$ in (f) uses $Ae^{-cL}$. Solid curves cover the fitted range and dashed curves show continuations. Open squares mark fitted intercepts or fitting-imposed zero limits. Parentheses denote errors estimated from fitting residuals.}
 \label{fig:spectra}
\end{figure*}

\section{Low-energy entanglement spectra across surface regimes}
\label{sec:spectra}

Figure~\ref{fig:spectra} shows the momentum-resolved entanglement spectra at $L=12$ and the finite-size evolution of the two gaps defined in Eqs.~\eqref{eq:gapk} and \eqref{eq:gap0}. The three surface regimes exhibit contrasting low-energy spectral structures, which we characterize through the relative evolution of these gaps and finite-size fits over $L=6,8,10,12$.

\subsection{Ordinary and special surfaces}
For the ordinary surface at $J_s=1.0$, both gaps decrease with increasing $L$ but remain finite over the sampled sizes [Fig.~\ref{fig:spectra}(d)]. We estimate their limiting values with
\begin{equation}
 \Delta(L)=\Delta_\infty+\frac{a}{L}+\frac{b}{L^2},
 \label{eq:finitegap}
\end{equation}
where $\Delta_\infty, a$ and $b$ are fitting parameters. The fits give $\Delta_{q_1,\infty}=2.644(9)$ and $\Delta_{0,\infty}=2.826(1)$. These positive extrapolations support a gapped interpretation in the probed sectors. They are notable because the ordinary physical surface still has critical correlations at the bulk critical point; that property need not imply a vanishing low-energy entanglement gap.

At $J_s=1.9895$, both gaps decrease more strongly [Fig.~\ref{fig:spectra}(e)]. Fits to $\Delta(L)=AL^{-p}$ give $p_{q_1}=0.496(3)$ and $p_0=0.4714(4)$. The nearest-momentum gap $\Delta_{q_1}$ remains approximately $2.8$ times the zero-momentum spacing $\Delta_0$ across all five displayed sizes. This nearly constant relative scale is direct evidence that the two excitations soften together over the available range, distinguishing the special spectrum from the finite scales observed at the ordinary surface. 

\subsection{Extraordinary surface and pair formation}
For the extraordinary surface at $J_s=3.0$, the two gaps exhibit markedly different finite-size behavior
[Fig.~\ref{fig:spectra}(f)]. The zero-momentum splitting decreases from $1.0082$ at $L=4$ to $0.05417$ at $L=12$, while the nearest-momentum excitation remains at a substantially higher energy. The ratio $\Delta_0/\Delta_{q_1}$ decreases from approximately $0.142$ to $0.00934$ over the same size range. This rapid reduction provides a direct finite-size signature of an increasingly near-degenerate zero-momentum pair, distinct from the comparable softening observed near the special coupling.

An exponential form, $\Delta_0(L)=Ae^{-cL}$, describes the observed splitting with $c=0.360(7)$, while fitting $\Delta_{q_1}$ to Eq.~\eqref{eq:finitegap} yields $\Delta_{q_1,\infty}=5.53(8)$. These empirical fits are consistent with a separation between the low-energy pair and the nearest-momentum excitation in the thermodynamic limit. 
Such a separation of scales is characteristic of finite-size Ising-ordered systems, where the mixing of opposite-magnetization configurations can produce a low-energy doublet~\cite{Pfeuty1970The}. The observed entanglement spectrum is consistent with this interpretation.

\subsection{Discussion of this section}
The results show that the spectral separation reveals the intrinsic mechanism of the surface criticality. The ordinary surface criticality in principle is totally decided by the bulk criticality, thus the entanglement spectrum exhibits a finite gap which means the low-energy behaviors come from the bulk critical modes. The extraordinary surface criticality induces long-range order on the boundary, meanwhile the entanglement spectra reflect the spontaneous symmetry breaking through the degeneracy of levels. The special point can be understood by a multi-criticality since the bulk critical modes couple with the gapless edge-modes together. Here we directly see the gapless edge-modes in the entanglement spectrum and the bulk critical modes modify its dispersion power. The coupled gapless modes lead to an effective long-range interaction for the entanglement Hamiltonian.

\section{Magnetic response of the surface entanglement Hamiltonian}
\label{sec:temperature}

To examine whether the corresponding reduced states exhibit magnetic responses compatible with the above spectral structures, we next turn to replica measurements of the second moment of magnetization Eq.~(\ref{eq:second-moment}) and the normalized Binder cumulant Eq.~(\ref{eq:binder}). 

\begin{figure}[tbp]
 \centering
 \includegraphics[width=\columnwidth]{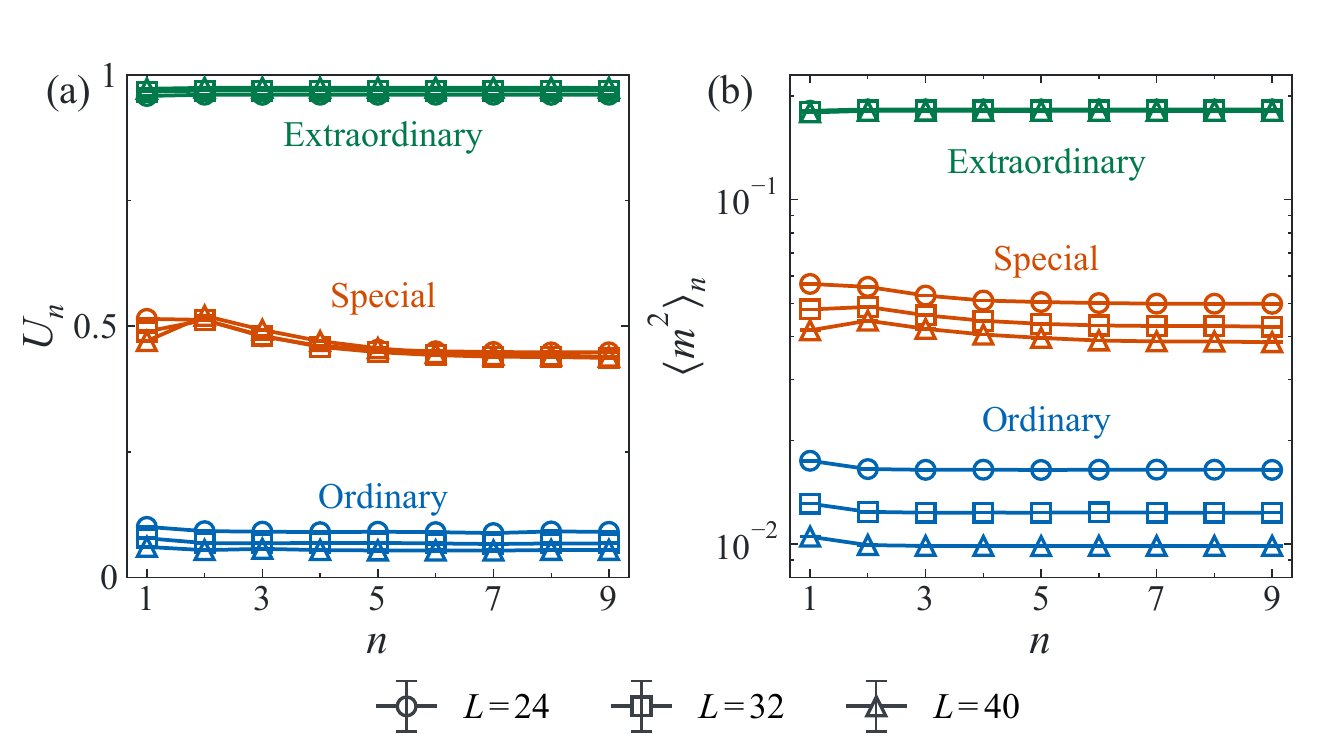}
 \caption{Replica magnetic response for reduced states constructed at $J_s=1.0$ (ordinary, blue), $1.9895$ (special, vermilion), and $3.0$ (extraordinary, green). (a) Normalized Binder cumulant $U_n$. (b) Surface magnetization second moment $\langle m^2\rangle_n$ on a logarithmic scale. Increasing $n$ lowers the entanglement temperature $T_E=1/n$ for each fixed $H_E$. Lines guide the eye.}
 \label{fig:replica}
\end{figure}

Figure~\ref{fig:replica} shows the replica magnetic response at the three representative physical surface couplings. The dependence on $n$ characterizes the response to decreasing $T_E$, while the system-size dependence helps assess whether ferromagnetic order persists as the system size increases.

For the ordinary surface at $J_s=1.0$, $U_n$ remains close to zero and decreases with increasing $L$, while $\langle m^2\rangle_n$ also becomes smaller for larger systems. Both observables vary weakly beyond the first few replicas. These trends are consistent with the absence of macroscopic ferromagnetic order over the sampled entanglement temperatures. 

For the extraordinary surface at $J_s=3.0$, $U_n$ stays close to unity, and $\langle m^2\rangle_n$ remains substantial with only weak size dependence. This ordering response is already present at $n=1$, where the replica ensemble reproduces the physical extraordinary surface, and persists over the sampled entanglement temperatures. Since both observables are invariant under spin reversal, they can reveal ferromagnetic ordering even when the average magnetization vanishes due to the $\mathbb{Z}_2$ symmetry. Together with the increasingly near-degenerate zero-momentum pair observed in Sec.~\ref{sec:spectra}, these results are consistent with an Ising-ordered interpretation of the low-energy entanglement structure.

At the special coupling $J_s=1.9895$, both observables exhibit a moderate dependence on $n$, while $\langle m^2\rangle_n$ decreases systematically with increasing $L$. Compared with the ordinary surface, the magnetic response is more pronounced, but it remains substantially weaker than that of the extraordinary surface, where $\langle m^2\rangle_n$ shows little size dependence. These results indicate that the special coupling does not exhibit the robust ferromagnetic ordering observed at $J_s=3.0$ over the sampled entanglement temperatures.

\begin{figure*}[tp]
 \centering
 \includegraphics[width=0.8\textwidth]{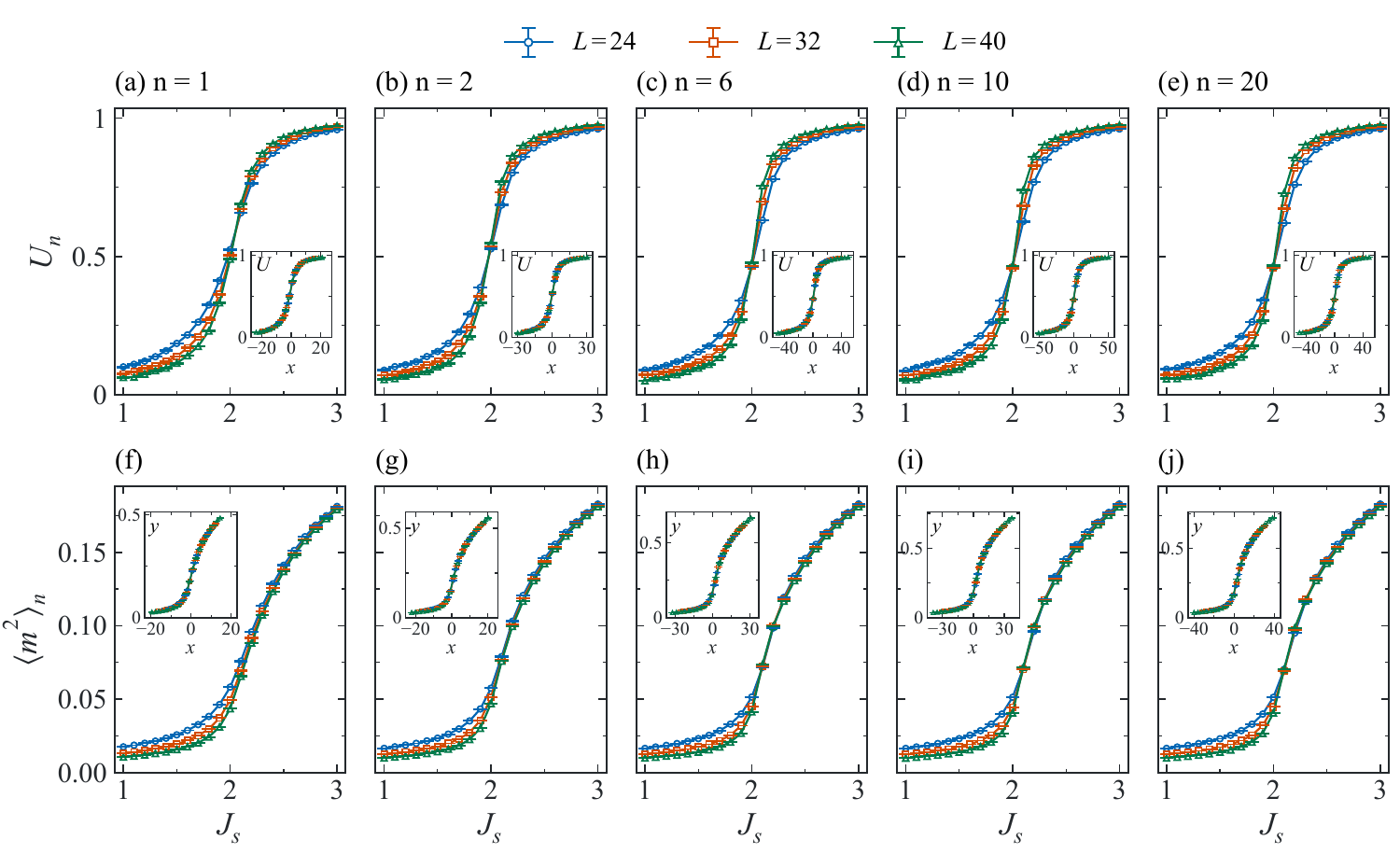}
 \caption{Surface-coupling scans for replica number $n=1,2,6,10,20$ (left to right), using $L=24,32,40$. (a)--(e) Binder cumulant $U_n$. (f)--(j) Surface second moment $\langle m^2\rangle_n$. Main panels show unscaled measurements. Insets retain the existing observable-specific collapses: $x=(J_s-J_{s,c}^{\rm fit})L^{1/\nu_{\rm fit}}$, with vertical rescaling $\langle m^2 \rangle_n L^{2\beta_{\rm fit}/\nu_{\rm fit}}$ for the second moment. Binder and second-moment fits use independent parameters. Points and error bars are means and standard errors over ten independent runs. Lines guide the eye.}
 \label{fig:coupling}
\end{figure*}

Additional surface-coupling scans for $n=1,2,6,10,20$, presented in Fig.~\ref{fig:coupling}, extend the magnetic analysis across the weak- and strong-coupling regimes. 
At weak surface coupling ($J_s\sim1.0$), the Binder cumulant $U_n$ is small and decreases toward zero with increasing system size, while the magnetization second moment $\langle m^2\rangle_n$ is progressively suppressed, consistent with a disordered regime. In contrast, at strong surface coupling ($J_s\sim3.0$), $U_n$ approaches unity and $\langle m^2\rangle_n$ exhibits much weaker size dependence, indicating a ferromagnetically ordered regime. 

Between these limits, the Binder curves for different system sizes exhibit crossings at intermediate couplings. Similar behavior is observed across the sampled replica numbers, suggesting the persistence of distinct disordered and ordered regimes as the entanglement temperature $T_E=1/n$ is varied. The estimated transition region 
changes only weakly for large replica numbers, suggesting an approximately temperature-independent ordering boundary at lower entanglement temperatures. This behavior is consistent with the schematic phase diagram in the $(J_s,T_E)$ plane shown in Fig.~\ref{fig:setup}(b).

To further characterize the finite-size behavior, we perform data collapses of $U_n$ and $\langle m^2\rangle_n$ at each sampled replica number assuming the scaling forms
\begin{align}
 U_n&=\mathcal U_n(x),\qquad
 \langle m^2\rangle_n=L^{-2\beta_{\rm fit}/\nu_{\rm fit}}\mathcal M_n(x),
 \label{eq:collapse}\\
 x&=(J_s-J_{s,c}^{\rm fit})L^{1/\nu_{\rm fit}}.\nonumber
\end{align}
The results are shown in the insets of Fig.~\ref{fig:coupling}.
Binder and second-moment parameters are fitted independently and Table~\ref{tab:collapse} retains the data from second-moment collapses. 
The physical $n=1$ ensemble supplies a benchmark: Tuning the surface enhancement at fixed bulk criticality gives $1/\nu_\text{fit}=0.715(1)$, i.e. $\nu_\text{fit}\approx 1.398$, and $\beta_\text{fit}=0.229(1)$~\cite{Deng2005Surface}. The deviations from the established surface critical exponents indicate that the fitted parameters should be interpreted as effective finite-size estimates rather than asymptotic critical exponents. 
 


\begin{table}[tp]
 \caption{Critical exponents obtained from the second-moment collapses in Fig.~\ref{fig:coupling}, using $L=24,32,40$. Parentheses give standard errors over ten bins.}
 \label{tab:collapse}
 \centering
 \begin{ruledtabular}
 \begin{tabular}{ccccc}
 $n$ & $J_{s,c}^{\rm fit}$ & $\beta_{\rm fit}$ & $\nu_{\rm fit}$ & $\beta_{\rm fit}/\nu_{\rm fit}$ \\
 1 & $2.144(5)$ & $0.175(2)$ & $1.30(2)$ & $0.135(3)$ \\
 2 & $2.046(3)$ & $0.183(1)$ & $1.21(1)$ & $0.152(2)$ \\
 6 & $2.021(2)$ & $0.1886(7)$ & $1.07(1)$ & $0.177(2)$ \\
 10 & $2.021(2)$ & $0.188(1)$ & $1.01(2)$ & $0.186(4)$ \\
 20 & $2.022(1)$ & $0.1877(7)$ & $1.006(7)$ & $0.187(2)$ \\
 \end{tabular}
 \end{ruledtabular}
\end{table}

\section{Constraints on the surface entanglement Hamiltonian}
\label{sec:longrange}

\subsection{Constraints from physical surface correlations}
\label{sec:constraints}

The exact relation between the physical surface and the entanglement ensemble at $T_E=1$ imposes constraints on the effective interactions of $H_E$. At the ordinary and special surface fixed points, the physical correlations decay algebraically~\cite{Diehl1997The,Przetakiewicz2025Boundary}, whereas in the extraordinary regime they approach a nonzero long-distance limit. 
The thermal state of $H_E$ at $T_E=1$ should reproduce these properties.

For a one-dimensional quantum spin Hamiltonian with translation-invariant, uniformly bounded, finite-range interactions, connected correlations decay exponentially at any positive temperature, and spontaneous Ising order is absent~\cite{Araki1969Gibbs,Bluhm2022Exponential}. Such a description is incompatible with the algebraic correlations of the ordinary and special surfaces or the long-range order of the extraordinary surface. Consequently, assuming a regular thermodynamic limit with uniformly bounded interaction strengths, the surface entanglement Hamiltonian cannot admit a fixed finite-range description in the physical surface-spin basis.

The argument does not determine the microscopic form or spatial decay of the effective interactions. In particular, it does not require a pairwise interaction or exclude multispin terms. Interactions extending over arbitrarily large distances provide a possible mechanism, consistent with previous studies of nonlocal entanglement Hamiltonians~\cite{Lundgren2013Entanglement,Luitz2014Participation,Li2024Relevant,Liu2026Worldline}. We illustrate this possibility using a long-range Ising chain, without assuming that it represents the exact entanglement Hamiltonian.

\subsection{Phenomenological comparison with a long-range Ising chain}

To illustrate an effective one-dimensional system that can accommodate finite-temperature magnetic order and the observed low-energy spectral structures, we consider a ferromagnetic transverse-field Ising chain with algebraically decaying interactions~\cite{Defenu2023Long-range},
\begin{align}
 H_{\rm LR}
 &=-\sum_{i<j}\frac{1}{r_{ij}^{\alpha}}\sigma_i^z\sigma_j^z
   -h\sum_i\sigma_i^x,
 \label{eq:hlr}\\
 r_{ij}
 &=\min\bigl(|i-j|,L-|i-j|\bigr).\nonumber
\end{align}
Here $r_{ij}$ is the shortest distance on a periodic chain. We choose $\alpha=2$ as a representative long-range interaction, for which finite-temperature ordering transitions are possible~\cite{Humeniuk2020Thermal}. This choice is phenomenological and is not inferred from the interaction kernel of $H_E$. The exchange amplitude is set to unity, each pair is counted once, and no Kac normalization is applied.

We consider the transverse fields $h_1=4.0$, $h_2=h_c=2.52253$, and $h_3=1.5$, representing the paramagnetic, critical, and ferromagnetic regimes, respectively. The thermodynamic critical field $h_c$ is taken from Ref.~\onlinecite{Koziol2021Quantum-critical}. We diagonalize the momentum sectors for $L=4,6,8,10,12$ and compare the same two gaps used in Sec.~\ref{sec:spectra}. 

\begin{figure*}[htp]
 \centering
 \includegraphics[width=0.8\textwidth]{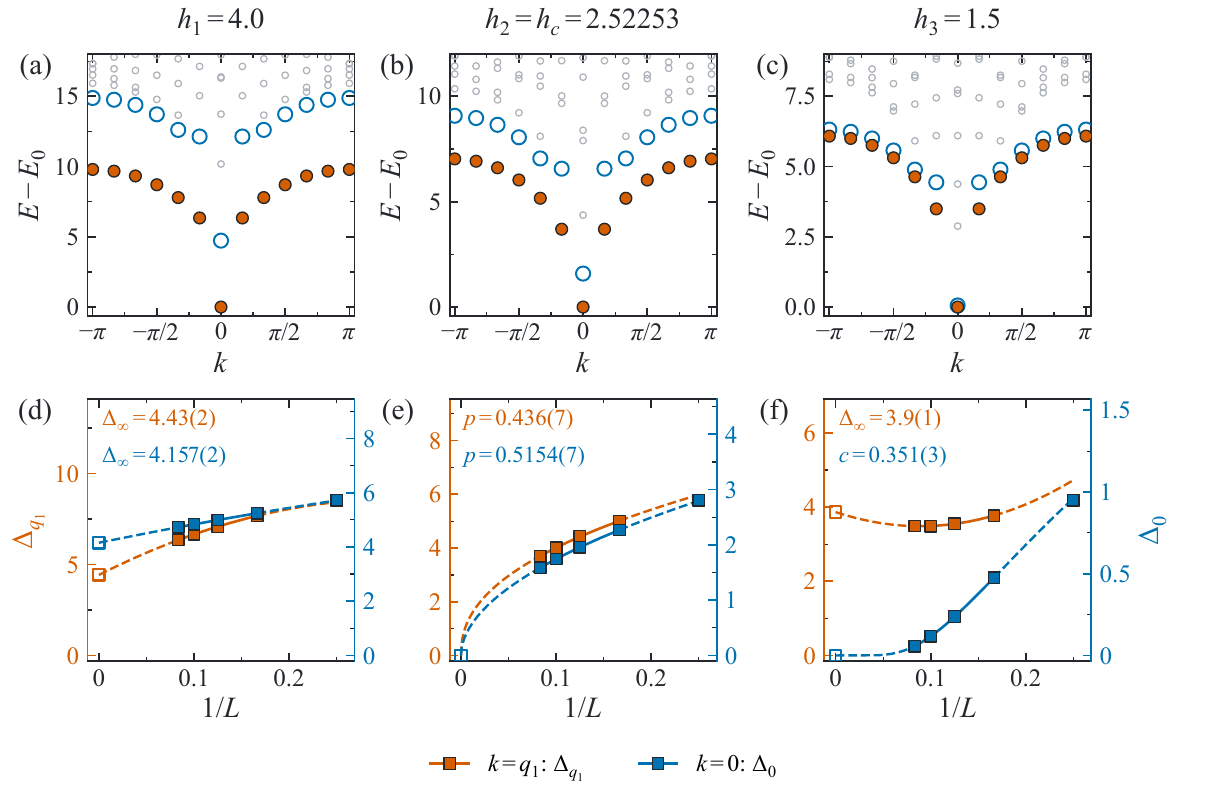}
 \caption{Spectra and finite-size gaps of the long-range Ising chain in Eq.~\eqref{eq:hlr}, used as a qualitative reference for the surface entanglement spectra. Columns correspond to the paramagnetic ($h_1=4.0$), critical ($h_2=h_c=2.52253$), and ferromagnetic ($h_3=1.5$) regimes. (a)--(c) Momentum-resolved energy spectra at $L=12$, shifted by $E_0$. Colors and markers follow Fig.~\ref{fig:spectra}. (d)--(f) Finite-size behavior of $\Delta_{q_1}=E_{\min}(q_1)-E_0(0)$ and $\Delta_0=E_1(0)-E_0(0)$. Fits use $L=6,8,10,12$, with $L=4$ displayed for comparison. Both gaps in (d) and $\Delta_{q_1}$ in (f) use Eq.~\eqref{eq:finitegap}; both gaps in (e) use $AL^{-p}$, and $\Delta_0$ in (f) uses $Ae^{-cL}$. Solid and dashed curves indicate the fitted ranges and their continuations, respectively. Open squares denote fitted intercepts or limits imposed by the fitting forms. Quoted uncertainties are estimated from fitting residuals.}
 \label{fig:longrange}
\end{figure*}

As shown in Fig.~\ref{fig:longrange}, the paramagnetic regime exhibits positive fitted gap intercepts, while both gaps decrease approximately algebraically at the critical field. In the ferromagnetic regime, the zero-momentum splitting decreases rapidly relative to the nearest-momentum excitation, producing an increasingly near-degenerate pair. These behaviors qualitatively reproduce the sequence of spectral structures observed across the ordinary, special, and extraordinary surfaces.

However, these spectral features are not unique to long-range Ising systems and therefore do not, by themselves, establish the nonlocality of $H_E$. The relevance of the long-range model is that this class of interactions can also support finite-temperature magnetic order, consistent with the constraint imposed by the physical extraordinary surface. The present comparison concerns the spectral organization and does not establish a quantitative correspondence between the thermal phase diagrams or critical exponents of the two systems. In particular, the inverse-square interaction is a representative realization rather than a reconstruction of the effective interactions in $H_E$.

\section{Discussion and Conclusion}
\label{sec:discussion}
Our results suggest that the low-energy entanglement spectrum provides a complementary way to characterize physical surface criticality. By keeping the bulk at the same quantum critical point and varying only the surface interactions, we observe distinct spectral structures associated with the ordinary, special, and extraordinary surface regimes. The retained subsystem is a complete physical surface chain, so its reduced density matrix encodes the influence of the critical bulk on the boundary degrees of freedom. In contrast to conventional characterizations based on surface correlation functions and scaling dimensions, momentum-resolved entanglement spectroscopy reveals how the dominant eigenstates of the reduced density matrix are organized in energy and momentum. This extends the connection between boundary critical phenomena and quantum entanglement to the low-energy structure of a surface entanglement Hamiltonian.

Near the special coupling, the zero-momentum splitting and the nearest-momentum excitation soften on comparable scales, whereas in the extraordinary regime the zero-momentum splitting decreases much more rapidly, producing an increasingly near-degenerate pair. The ordinary regime instead exhibits positive extrapolated gaps in both probed sectors. These contrasting behaviors show how the relative evolution of momentum-resolved excitation scales can distinguish critical softening from the formation of an ordered-state doublet. They provide candidate spectral signatures of the underlying surface universality classes, although establishing their asymptotic universality requires further finite-size analysis.

The replica magnetic measurements in entanglement ensemble provide an additional perspective on these spectral structures. The robust magnetic response in the extraordinary regime is consistent with an Ising-ordered interpretation of the low-energy doublet, while the ordinary and special surfaces exhibit substantially weaker ordering tendencies over the sampled entanglement temperatures. Nevertheless, the entanglement spectrum and replica observables probe different aspects of the same reduced state: the former determines the excitation energies, whereas the latter also depends on the magnetic properties of the corresponding states. In particular, the ordinary physical surface exhibits algebraically decaying correlations at $T_E=1$, despite the positive extrapolated gaps in the probed low-energy sectors. This illustrates that physical boundary criticality does not directly determine the low-temperature spectral properties of the entanglement Hamiltonian. 

A further implication concerns the effective interactions of the surface entanglement Hamiltonian. Although the three surface regimes exhibit different low-energy spectral structures, their physical correlations impose a common constraint: the algebraic correlations of the ordinary and special surfaces and the long-range order of the extraordinary surface are incompatible with a regular one-dimensional Hamiltonian with uniformly bounded, fixed finite-range interactions. Thus, the locality properties of the entanglement Hamiltonian alone do not determine its low-energy spectral organization. The long-range Ising chain illustrates how nonlocal interactions can accommodate similar spectral structures and finite-temperature ordering, without establishing a microscopic correspondence between the two Hamiltonians. More generally, our results point to the possibility that boundary critical regimes organize distinct low-energy entanglement structures even when the associated entanglement Hamiltonians share the same locality constraint. Establishing which features are universal, and how they relate quantitatively to boundary critical data, remains an important direction for future work.

In summary, we have investigated the low-energy entanglement spectrum of a single physical surface of the square-lattice transverse-field Ising model at a fixed bulk quantum critical point. Momentum-resolved spectra reveal distinct finite-size signatures across the ordinary, special, and extraordinary surface regimes: positive extrapolated gaps in the probed ordinary sectors, comparable softening of low-energy scales near the special coupling, and an increasingly near-degenerate zero-momentum pair in the extraordinary regime. Replica magnetic measurements provide complementary ordering information, while the physical surface correlations at $T_E=1$ suggest the nonlocality of the corresponding entanglement Hamiltonian. Together, these findings demonstrate how different physical surface critical regimes can be distinguished through their low-energy entanglement spectral structures.

\vspace*{0.5cm}
\section*{acknowledgements}
We thank Junchen Rong for helpful discussions. B.-B. M. was supported by  the Natural Science Foundation of Shandong Province, China (Grant No. ZR2024QA194). Z. W. was supported by the National Natural Science Foundation of China under Grants No. 12605078.
This project is supported by the Scientific Research Project (No. WU2025B011), Feng-Ying Career Development Chair Fund and the Start-up Funding of Westlake University, and the Natural Science Foundation of China (Grant No. 12674188), and Quantum MetaKey (QuaMeK) computing. The authors thank the IT service office and the high-performance computing center of Westlake University.

\bibliographystyle{apsrev4-2}
\bibliography{references}
\end{document}